# Adiabatic decay of internal solitons due to Earth' rotation within the framework of the Gardner–Ostrovsky equation

Maria Obregon[1], Nawin Raj[2], Yury Stepanyants[2,3*)]

[1] E.T.S. Ingeniería Industrial, University of Malaga,
Dr Ortiz Ramos s/n, 29071, Malaga, Spain;
[2] Faculty of Health, Engineering and Sciences, University of Southern Queensland,
West St., Toowoomba, QLD, 4350, Australia;
[3] Department of Applied Mathematics, Nizhny Novgorod State Technical University
n.a. R.E. Alekseev, Nizhny Novgorod, 603950, Russia.

## Abstract

The adiabatic decay of different types of internal wave solitons caused by the Earth' rotation is studied within the framework of the Gardner–Ostrovsky equation. The governing equation describing such processes includes quadratic and cubic nonlinear terms, as well as the Boussinesq and Coriolis dispersions: $(u_t + c\,u_x + \alpha\,u\,u_x + \alpha_1\,u^2\,u_x + \beta\,u_{xxx})_x = \gamma u$. It is shown that at the early stage of evolution solitons gradually decay under the influence of weak Earth' rotation described by the parameter $\gamma$. The characteristic decay time is derived for different types of solitons for positive and negative coefficient of cubic nonlinearity $\alpha_1$ (both signs of that parameter may occur in the oceans). The coefficient of quadratic nonlinearity $\alpha$ determines only a polarity of solitary wave when $\alpha_1 < 0$ or the asymmetry of solitary waves of opposite polarity when $\alpha_1 > 0$. It is found that the adiabatic theory describes well the decay of solitons having bell-shaped profiles. In contrast to that, large amplitude table-top solitons, which can exist when $\alpha_1$ is negative, are structurally unstable. Under the influence of Earth' rotation they transfer first to the bell-shaped solitons, which decay then adiabatically. Estimates of the characteristic decay time of internal solitons are presented for the real oceanographic conditions.

---

[*)] Corresponding author: Yury.Stepanyants@usq.edu.au

## 1. Introduction

The model Gardner–Ostrovsky (GO) equation, was derived for the description of long internal waves of large amplitude (Holloway et al., 1999):

$$\frac{\partial}{\partial x}\left(\frac{\partial u}{\partial t} + c\frac{\partial u}{\partial x} + \alpha u \frac{\partial u}{\partial x} + \alpha_1 u^2 \frac{\partial u}{\partial x} + \beta \frac{\partial^3 u}{\partial x^3}\right) = \gamma u, \qquad (1.1)$$

where $c$ is the speed of dispersionless linear waves, $\alpha$ and $\alpha_1$ are the coefficient of quadratic and cubic nonlinearities, respectively, and $\beta$ and $\gamma$ are the coefficients of small-scale (Boussinesq) and large-scale (Coriolis) dispersions, respectively. The variable $u(x, t)$ describes a perturbation of an isopycnal surface (the surface of equal density) from its rest position. For internal waves in two-layer fluid in the Boussinesq approximation the parameters of GO equation are (Apel et al., 2007):

$$c = \sqrt{\frac{\delta\rho}{\rho_m} g \frac{h_1 h_2}{h_1 + h_2}}; \quad \alpha = \frac{3}{2} c \frac{h_1 - h_2}{h_1 h_2}; \quad \alpha_1 = -\frac{3}{8} c \frac{(h_1 + h_2)^2 + 4 h_1 h_2}{(h_1 h_2)^2}; \quad \beta = \frac{c}{6} h_1 h_2; \quad \gamma = \frac{f^2}{2c}, \qquad (1.2)$$

where $\delta\rho$ is the density difference between the lower and upper layers, $\rho_m$ is the mean water density, $g$ is the acceleration due to gravity, $h_1$ and $h_2$ are thicknesses of upper and lower layers, respectively, and $f$ is the Coriolis parameter characterizing the Earth' rotation at the particular geographical latitude.

Equation (1.1) combines the dispersion effects due to non-hydrostaticity caused by the finiteness of basin depth (the Boussinesq dispersion proportional to $\beta$) and due to Earth's rotation (the Coriolis dispersion proportional to $\gamma$). The equation contains also two nonlinear terms proportional to $\alpha$ and $\alpha_1$. The former one is the traditional quadratic nonlinear term appearing due to hydrodynamic nonlinearity, as in the Korteweg–de Vries (KdV) equation (Whitham, 1974), whereas the latter term appears either when the first term becomes anomalously small [such situation arises in the internal wave dynamics (Holloway et al., 1999)], or when the GO equation is used as the model equation to describe approximately large-amplitude waves (Michallet & Barthélemy, 1998; Ostrovsky & Stepanyants, 2005; Apel et al., 2007).

Equation (1.1) is apparently non-integrable and even its stationary solutions are unknown. Moreover, according to "antisoliton theorem", stationary solitary type solutions of this equation are impossible if $\beta\gamma > 0$ (Leonov, 1981; Galkin & Stepanyants, 1991). In the meantime, in the absence of rotation ($\gamma = 0$) the GO equation reduces to the well-known and completely integrable Gardner equation (Slyunyaev & Pelinovsky, 1999; Slyunyaev, 2001). The latter equation has soliton solutions whose profile essentially depends on the amplitude and sign of cubic coefficient $\alpha_1$, whereas the coefficient $\alpha$ determines only a solitary wave polarity when $\alpha_1 < 0$ (a soliton has positive polarity (hump wave) if $\alpha > 0$, and negative polarity (depression wave) otherwise) or the asymmetry of solitary waves of opposite polarity when $\alpha_1 > 0$ (see below for clarification). It is a matter of interest to study the influence of weak rotation on the dynamics of quasi-stationary Gardner solitons in application to large amplitude internal waves. Such waves are often observed in shallow coastal regions where they may have an influence on human activity, engineering constructions, off-shore petroleum exploration, production and sub-sea storage activities, etc.

In another limiting case of very long internal waves the small-scale Boussinesq dispersion can be neglected, then equation (1.1) reduces to the equation which is knowns as the reduced Gardner–Ostrovsky (rGO) equation (Obregon & Stepanyants, 2014):

$$\frac{\partial}{\partial x}\left(\frac{\partial u}{\partial t} + c\frac{\partial u}{\partial x} + \alpha u\frac{\partial u}{\partial x} + \alpha_1 u^2 \frac{\partial u}{\partial x}\right) = \gamma u. \tag{1.3}$$

The particular versions of this equation were considered in many papers starting from the original paper by Ostrovsky (Ostrovsky, 1978) [see also the reviews (Ostrovsky & Stepanyants, 1990; Grimshaw et al., 1998b; Stepanyants, 2006) and references therein].

In this paper we present asymptotic solutions for slowly varying Gardner solitons of internal waves due to influence of Earth' rotation. We show that the rotation leads to soliton terminal decay; we estimate the life time of Gardner solitons and characteristic spatial scales of their decay. It is found that the character of soliton decay is different for the bell-shaped and table-top solitons – in the former case solitons decay adiabatically keeping their profiles, whereas in the latter case, solitons, being structurally unstable, quickly transfer first into the bell-shaped solitons which decay then adiabatically.

To study the process of soliton decay it is convenient to transfer first Eq. (1.1) into the dimensionless form using the normalized variables:

$$\xi = \frac{x-ct}{L_0}; \quad \tau = t\frac{\alpha U_0}{L_0}; \quad \upsilon = \frac{u}{U_0}, \tag{1.4}$$

where $U_0$ is the characteristic amplitude of initial perturbation, and $L_0$ is its characteristic width (these parameters will be specified below). In the new variables Eq. (1.1) reads:

$$\frac{\partial}{\partial \xi}\left(\frac{\partial \upsilon}{\partial \tau} + \upsilon\frac{\partial \upsilon}{\partial \xi} + \mu\upsilon^2\frac{\partial \upsilon}{\partial \xi} + \frac{1}{\mathrm{Ur}}\frac{\partial^3 \upsilon}{\partial \xi^3}\right) = \varepsilon\upsilon, \tag{1.5}$$

where $\mu = \alpha_1 U_0/\alpha$, $\varepsilon = \gamma L_0^2/(\alpha U_0)$, and $\mathrm{Ur} = \alpha U_0 L_0^2/\beta$ is the well-known Ursell parameter in the theory of shallow water waves [see, e.g., (Whitham, 1974; Dingemans, 1997)].

The pulse-type initial condition for Eq. (1.5) reads: $\upsilon(0, \xi) = F(\xi)$, where function $F(\xi)$ has a unit amplitude and unite width.

In the typical oceanic conditions the coefficients of Eq. (1.1) are such that $\alpha < 0$, $\beta > 0$, $\gamma > 0$, whereas the coefficient $\alpha_1$ may be both positive and negative (distributions of all these coefficients in the World Ocean are presented in (Grimshaw et al., 2007)). As the consequence of that, the dimensionless coefficient $\mu$ may be also both positive and negative, whereas the parameters Ur and $\varepsilon$ in Eq. (1.5) are always positive for oceanic waves.

## 2. Influence of rotation on the dynamics of Gardner solitons when $\alpha_1 < 0$

### 2.1 The theoretical analysis

We start our analysis with the most typical oceanic case when $\alpha_1$ is negative (and hence $\mu$ is negative too). Presume also that the initial perturbation is a soliton whose polarity is negative in this case (Apel et al., 2007) and present Eq. (1.5) in the form which is more convenient for the application of the perturbation analysis, namely

$$\frac{\partial \upsilon}{\partial \tau} + \upsilon \frac{\partial \upsilon}{\partial \xi} + \mu \upsilon^2 \frac{\partial \upsilon}{\partial \xi} + \frac{1}{\mathrm{Ur}} \frac{\partial^3 \upsilon}{\partial \xi^3} = -\varepsilon \int_{\xi}^{+\infty} \upsilon(\xi') d\xi', \qquad (2.1)$$

where limits of integration in the right-hand side are chosen such that the perturbation is zero far away from the soliton front, i.e. at $\xi = \infty$.

When $\varepsilon = 0$, Eq. (2.1) reduces to the well-known Gardner equation (alias extended or combined KdV equation). One of the exact stationary solutions to this equation is the soliton which can be presented in different equivalent forms [see, e.g., (Apel et al., 2007; Grimshaw et al., 2003; 2010; Ostrovsky et al., 2015)]; here we will use the following two forms:

$$\upsilon = \frac{A}{1 + B \cosh\left(\frac{\xi - V\tau}{\Delta/2}\right)}, \qquad (2.2a)$$

where $0 \leq B \leq 1$ and all other parameters can be presented in terms of $B$:

$$A = \frac{1-B^2}{-\mu}, \quad \Delta = \sqrt{\frac{-24\mu}{\mathrm{Ur}(1-B^2)}}, \quad V = \frac{1-B^2}{-6\mu}. \qquad (2.3a)$$

We will be using also another form of soliton solution which is equ1ivalent to Eq. (2.2a) but reads differently:

$$\upsilon = -\frac{v}{2\mu}\left[\tanh\left(\frac{\xi - V\tau}{\Delta} + \phi\right) - \tanh\left(\frac{\xi - V\tau}{\Delta} - \phi\right)\right], \qquad (2.2b)$$

where $v^2 = 1 - B^2$ and other three parameters can be presented in terms of $v$:

$$\phi(v) = \frac{1}{4} \ln\left(\frac{1+v}{1-v}\right), \quad \Delta(v) = \frac{2L_0}{v} \sqrt{-\frac{6\mu}{\text{Ur}}}, \quad V(v) = -\frac{v^2}{6\mu}. \tag{2.3b}$$

The amplitude of Gardner soliton (2.2a) is determined by the formula

$$U = \frac{A}{1+B} \equiv -\frac{1-B}{\mu}. \tag{2.4}$$

The soliton profile varies with the parameter $B$ from the bell-shaped KdV soliton, when $B \to 1$, to the table-top soliton, when $B \to 0$. Figure 1 shows soliton solution (2.2) for several values of parameter $B$. The table-top soliton resembles a meander-type pulse, which can be treated as a pair of stationary moving kink (dissipationless shock wave) and anti-kink (see line 3 in Fig. 1). The kink is described by the first tanh-function in Eq. (2.2b), whereas the anti-kink is described by the second tanh-function in that equation. Such representation is especially helpful when the widths of kink and anti-kink fronts $\Delta$ are much less than the distance between them.

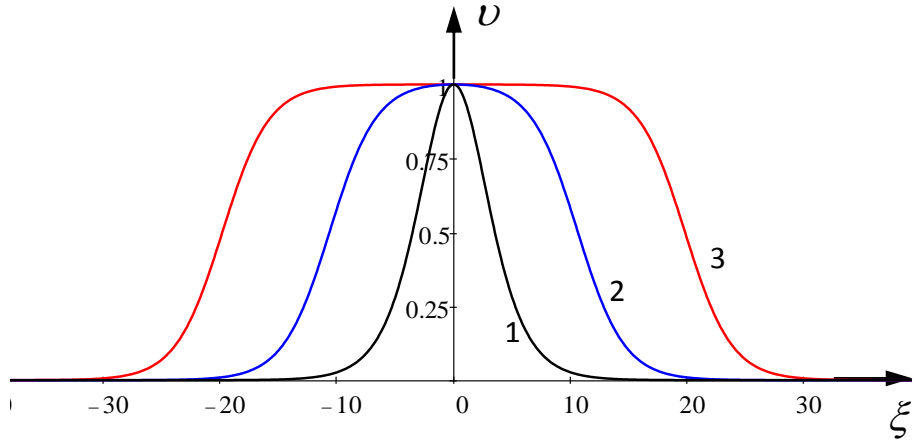

Fig. 1. Normalised Gardner soliton (2.2a), for several values of parameter $B$. Line 1: $B = 0.9999$ (KdV soliton); line 2: $B = 10^{-2}$ ("fat soliton"); line 3: $B = 10^{-4}$ (table-top soliton).

In the KdV limit ($B \to 1$, $v \to 0$) solution (2.2) reduces to

$$\upsilon = \frac{A}{1 + B\cosh\left(\frac{\xi - V\tau}{\Delta/2}\right)} \to \frac{A}{2\cosh^2\left(\frac{\xi - V\tau}{\Delta}\right)}. \tag{2.5}$$

If $\varepsilon \neq 0$, but sufficiently small, $\varepsilon \ll 1$, then solution (2.2) is no longer valid. However, if the Gardner soliton is structurally stable, then under the influence of small perturbative term in the right-hand side of Eq. (2.1) it may experience just a gradual adiabatic variation with time, keeping the profile which corresponds to the instant value of the parameter $B(\tau)$ and the relationships between other parameters (amplitude, velocity, width) at any instant of time. Then the evolution of soliton parameters with time can be evaluated with the help of perturbation theory developed in many papers [see, for instance, Ref. (Grimshaw et al., 1998a; 2003; 2010) in application to waves in a rotating fluid described by the Ostrovsky equation]. The application of perturbation theory reduces, in essence, to the energy balance equation for a soliton. Such approach was successfully used for the calculation of adiabatic evolution of Gardner solitons in viscous fluids with different mechanisms of dissipation (Grimshaw et al., 2003; 2010; 2018; Clarke et al., 2018). Here we apply a similar approach to calculate the time variation of soliton parameters under the influence of fluid rotation.

Multiplying Eq. (2.1) by $\upsilon$ and integrating then the resultant equation over $\xi$ in the infinite limits we obtain:

$$\frac{d}{d\tau}\int_{-\infty}^{+\infty}\upsilon^2 d\xi = -\varepsilon\left[\int_{-\infty}^{+\infty}\upsilon(\xi)d\xi\right]^2. \tag{2.6}$$

Substitute now into this equation soliton solution (2.2a) assuming that the parameter $B$ is a slowly varying function of time. After integration and simple manipulations we obtain

$$\frac{dB}{d\tau} = \varepsilon\sqrt{\frac{-3\mu}{2\mathrm{Ur}}}\frac{B}{\sqrt{1-B^2}}\ln^2\frac{1-\sqrt{1-B^2}}{1+\sqrt{1-B^2}}. \tag{2.7}$$

In general, an analytical solution to this equation can be presented in the implicit form in a quadrature:

$$\varepsilon\tau\sqrt{\frac{-3\mu}{2\mathrm{Ur}}} = \int_{B_0}^{B}\frac{\sqrt{1-B^2}\,dB}{B\left[\ln\left(1-\sqrt{1-B^2}\right)-\ln\left(1+\sqrt{1-B^2}\right)\right]^2}, \tag{2.8}$$

where $B_0$ is the initial value of parameter $B$ at $\tau = 0$.

In the KdV limit ($B_0 \to 1$, $\mu \sim B_0 - 1 \to 0$, $\text{Ur} \to 12$), Eq. (2.7) simplifies and reads

$$\frac{dB}{d\tau} = 2\varepsilon\sqrt{(1-B_0)(1-B)}. \qquad (2.9)$$

Solution to this equation can be readily found in the explicit form:

$$B = 1 - (1 - B_0)(1 - \varepsilon\tau)^2. \qquad (2.10)$$

This solution can be presented in terms of soliton amplitude [see, e.g., (Grimshaw et al., 1998a; 1998b; Fraunie & Stepanyants, 2002)]:

$$\frac{U}{U_0} = (1 - \varepsilon\tau)^2. \qquad (2.11)$$

As follows from this formula, a soliton completely vanishes in a finite time $\tau = \tau_{ext} \equiv 1/\varepsilon$. But in fact it transfers asymptotically after long-time evolution into an envelope soliton and non-stationary dispersive wave train. This was revealed for the first time in Ref. (Helfrich, 2007) within the framework of fully nonlinear set of Boussinesq equations (augmented by the Coriolis force) for internal waves in two-layer rotating fluid, and then studied in detail both within the framework of Ostrovsky equation (Grimshaw & Helfrich, 2008; 2012; Grimshaw et al., 2013; 2016; Whitfield & Johnson, 2014) and within the framework of Gardner–Ostrovsky equation (Whitfield & Johnson, 2015). The envelope soliton is described by the generalized non-linear Schrödinger equation, and its carrier wavenumber $k_c = (3\varepsilon\text{Ur}/4)^{3/4}$ is close to the maximum of growth rate of modulation instability.

In another limit, when $B_0 \to 0$ ($\mu \to -1$, $\text{Ur} \to 24$), Eq. (2.7) again simplifies and reduces to

$$\frac{dB}{d\tau} = \varepsilon\sqrt{\frac{-24\mu}{\text{Ur}}} B \ln^2 \frac{B}{2}. \qquad (2.12)$$

It can be explicitly integrated resulting in

$$B = 2\exp\left[\frac{\ln(B_0/2)}{1 - \varepsilon\tau\ln(B_0/2)\sqrt{-24\mu/\text{Ur}}}\right]. \qquad (2.13)$$

In terms of soliton amplitude this gives:

$$U(\xi) = \frac{1}{-\mu}\left\{1 - 2\exp\left[\frac{\ln(B_0/2)}{1 - \varepsilon\tau \ln(B_0/2)\sqrt{-24\mu/\mathrm{Ur}}}\right]\right\} \xrightarrow[B_0 \to 0]{} 1 - 2\exp\left(-\frac{1}{\varepsilon\tau}\right). \quad (2.14)$$

According to this formula, soliton amplitude turns to zero at a finite time, in particular, for large amplitude soliton with $B_0 \to 0$ the extinction time is $\tau_{ext} \equiv 1/(\varepsilon \ln 2)$. Hence, the extinction time of the table-top soliton is greater than the extinction time of KdV soliton by the factor of $1/\ln 2 \approx 1.443$.

In general, Eq. (2.7) can be easily solved numerically and after that, using the relationships between the parameter $B$ and other soliton parameters as per Eqs. (2.3a) and (2.4), one can find time dependence of soliton amplitude $U(\tau)$, velocity $V(\tau)$, front width $\Delta(\tau)$, and the total soliton width $D(\tau)$. Time dependence of soliton amplitude is shown in Fig. 2 for different initial values of parameter $B$.

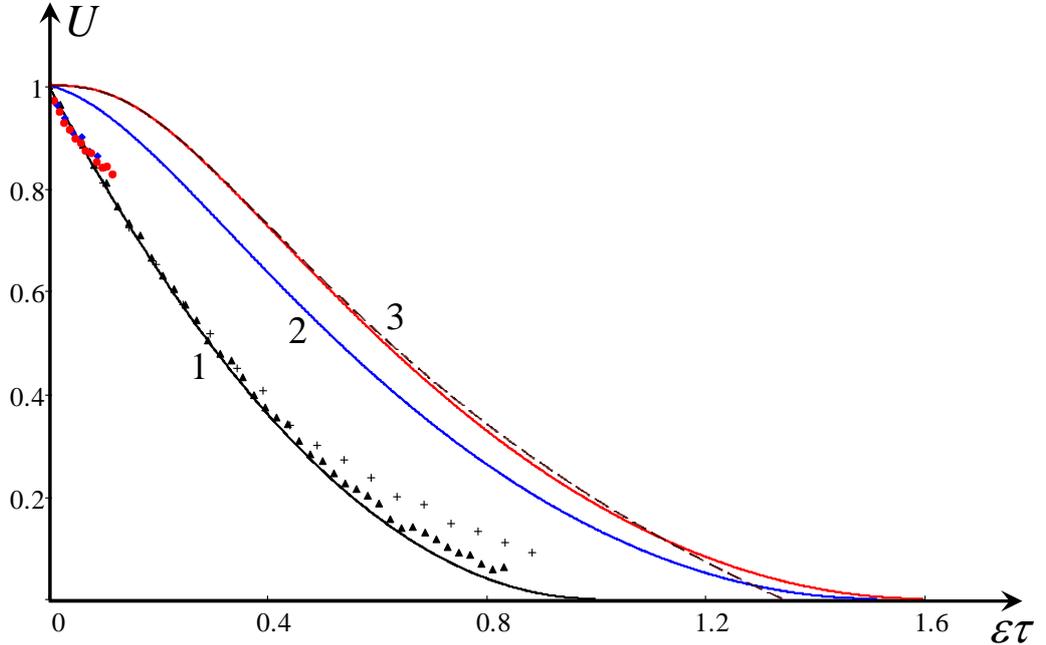

Fig. 2. Soliton amplitude against time in normalized variables. Line 1: $B_0 = 0.9999$ (KdV soliton); line 2: $B_0 = 10^{-2}$ ("fat soliton"); solid line 3: $B_0 = 10^{-4}$ (table-top soliton). Dashed line next to line 3 represents the asymptotic dependence (2.14). Symbols represent numerical data: triangles for $B_0 = 0.9999$, pluses – for $B_0 = 10^{-2}$, and dots – for $B_0 = 10^{-4}$.

Asymptotic dependence (2.11) for the KdV soliton completely coincides with the numerical solution shown in Fig. 2 by line 1. Another asymptotic solution (2.14) corresponding to the case of small $B_0 = 10^{-4}$ is shown by the dashed line next to the solid line 3 which was obtained by numerical solution of Eq. (2.7). As one can see, there is a good agreement between the asymptotic and numerical solutions of Eq. (2.7). In all cases soliton amplitudes monotonically decrease with time independently of initial value of the governing parameter $B$. At a certain time, the amplitude formally vanishes within the framework of the adiabatic theory. The corresponding extinction time has been presented above for two limiting cases of KdV soliton ($B_0 \to 1$) and table-top soliton ($B_0 \to 0$). In general, the extinction time can be found from Eq. (2.8) when $B$ turns to unity; then we have:

$$\varepsilon \tau_{ext}(B_0) = \frac{4}{\sqrt{1-B_0^2}} \int_{B_0}^{1} \frac{\sqrt{1-B^2}\, dB}{B\left[\ln\left(1-\sqrt{1-B^2}\right) - \ln\left(1+\sqrt{1-B^2}\right)\right]^2}. \tag{2.15}$$

Figure 3 shows the dependence of normalized extinction time on $B_0$ as per Eq. (2.15). As follows from Eq. (2.15), the extinction time abruptly goes to infinity in the limit of a very wide table-top soliton, when $B_0 \to 0$, whereas the approximate formula (2.14) provides a reasonable estimate for the extinction time $\varepsilon \tau_{ext} = 1/\ln 2 \approx 1.443$.

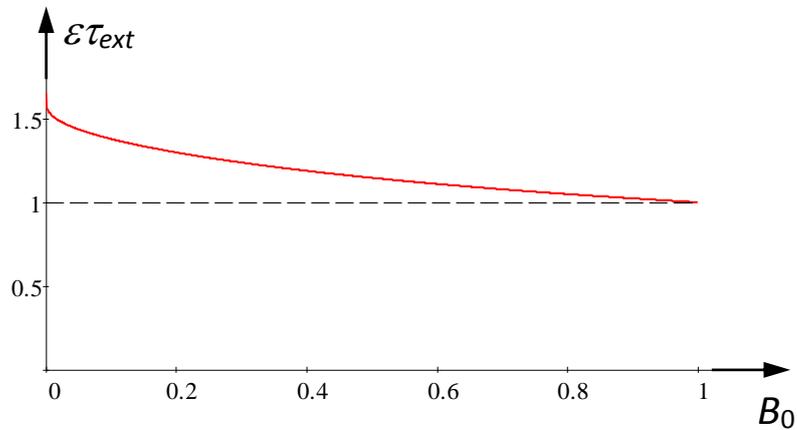

Fig. 3. The extinction time of the Gardner soliton against the initial value of the parameter $B$.

The minimum value of the extinction time realizes for the KdV soliton: $\tau_{ext} = 1/\varepsilon$. In the dimensional variables this gives $T_{ext} = 1/(\gamma L_0) = (1/\gamma)(\alpha U_0/12\beta)^{1/2}$. Thus, the extinction time depends not only on the parameters of the GO equation, but also on the initial soliton amplitude $U_0$ too. Using the coefficients (1.1) and taking the following set of parameters:

$$\delta\rho/\rho_m = 5\cdot 10^{-4}, f = 10^{-4} \text{ s}^{-1}, h_1 = 24 \text{ m}, h_2 = 26 \text{ m},$$

we obtain for the initial soliton of amplitude $U_0 = 2$ m:

$$T_{ext} = \frac{1}{f^2}\sqrt{\frac{\delta\rho}{\rho_m} g \frac{3U_0(h_1 - h_2)}{h_1 h_2 (h_1 + h_2)}} \approx 1.39 \cdot 10^5 \text{ s} \approx 38.6 \text{ h}.$$

(Note that soliton polarity is such that the product $U_0(h_1 - h_2)$ is always positive.) The extinction time can be compared with the total soliton duration which can estimated as:

$$T_t = \frac{2\Delta}{c} \equiv \frac{2}{c}\sqrt{\frac{12\beta}{\alpha U_0}} = 4\sqrt{\frac{h_1 h_2 (h_1 + h_2)}{3U_0(h_1 - h_2)g(\delta\rho/\rho_m)}} \approx 2.88 \cdot 10^3 \text{ s} \approx 0.8 \text{ h}.$$

Thus under the chosen set of parameters, the extinction time of the KdV soliton due to Earth' rotation is about 50 times greater than its characteristic duration.

The soliton velocity is related to the amplitude; the relationship between them can be derived from Eqs. (2.3a) and (2.4):

$$V = \frac{U}{3}\left(1 + \frac{\mu U}{2}\right). \tag{2.16}$$

In the adiabatic approximation, time dependence of soliton velocity follows the variation of amplitude in accordance with Eq. (2.16). The dependence of $V(\tau)$ is shown in Fig. 4 for the same three initial values of the parameter $B$ as in Fig. 2.

The soliton velocity also monotonically decrease with time independently of the initial value of the governing parameter $B$. The traversed path for the KdV soliton until its disappearance can be easily calculated:

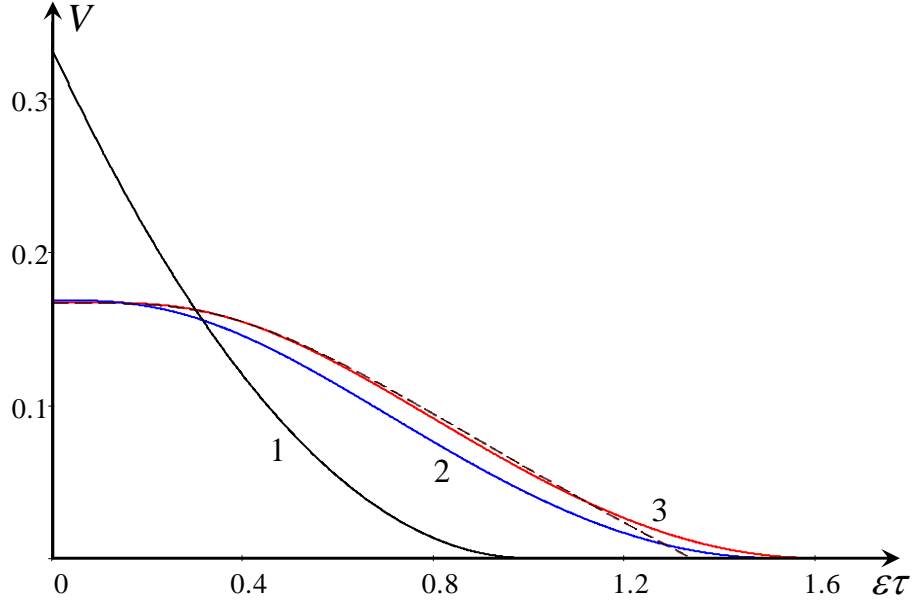

Fig. 4. Soliton velocity against time in normalized variables. Line 1: $B_0 = 0.9999$ (KdV soliton); line 2: $B_0 = 10^{-2}$ ("fat soliton"); solid line 3: $B_0 = 10^{-4}$ (table-top soliton). Dashed line next to line 3 corresponds to the asymptotic dependence (2.14).

$$\frac{S_{KdV}}{L_0} = \int_0^{\tau_{ext}} V(\tau)d\tau = \frac{1}{3}\int_0^{\tau_{ext}} \frac{U(\tau)}{U_0}d\tau = \frac{1}{3}\int_0^{\tau_{ext}} (1-\varepsilon\tau)^2\,d\tau = \frac{1}{9\varepsilon}. \qquad (2.17)$$

In the case of the table-top soliton with $B_0 = 10^{-4}$ the total traversed path can be evaluated numerically; the result is $S_{tts}/L_0 \approx 0.142/\varepsilon$. For the limiting case of the table-top soliton with $B_0 = 0$ the total traversed path can be calculated analytically using Eqs. (2.14) and (2.16); the result is $S_{lim}/L_0 = -4\mathrm{Ei}(-\ln 4)/3\varepsilon \approx 0.159/\varepsilon$, where $\mathrm{Ei}(x)$ is the exponential integral function of $x$.

The characteristic soliton scale (the width of the soliton front) $\Delta(\tau)$ and the total soliton width can be also readily found in terms of parameter $B(\tau)$. The total soliton width can be defined as the distance between the soliton front and rear slope at the level of half of soliton amplitude, i.e. when $\upsilon(D, \tau) = U/2$ for any instant of time [see (Apel et al., 2007)]. Using Eq. (2.2b) and the relationship between $v$ and $B$, $v^2 = 1 - B^2$, one can derive

$$\frac{D}{L_0} = \sqrt{-\frac{6\mu}{\mathrm{Ur}}}\frac{4}{\sqrt{1-B^2}}\ln\left[\frac{\sqrt{1+\sqrt{1-B^2}} + \sqrt{1-\sqrt{1-B^2}} + \sqrt{2(1+3B)}}{2\sqrt{B}}\right]. \qquad (2.18)$$

Graphics of soliton front width $\Delta(\tau)$ and the total soliton width $D(\tau)$ against time are shown in Fig. 5 for different values of $B_0$. In the course of soliton propagation, the soliton front width $\Delta(\tau)$ monotonically increases with time, whereas the dependence of total soliton width $D(\tau)$ may be non-monotonic depending on the initial value of parameter $B$. The minimum value of $D(\tau)$ occurs at $B \approx 0.451$ and equals to $D_{min} = 4.746 \cdot (-6\mu/\text{Ur})^{1/2}$. This value is attained at a certain instant of time, if the initial soliton amplitude is large enough, i.e. if $B_0 < 0.451$. Thus, a small-amplitude soliton with $B_0 > 0.451$, whose initial width $D > D_{min}$, decays in time in the course of propagation, and its width monotonically increases with time, whereas large-amplitude soliton with $B_0 < 0.451$, whose initial width $D$ is also greater than $D_{min}$, shrinks in time first, attains the minimal value $D_{min}$, and after that expands with time decreasing in amplitude. This is illustrated, for example, by solid lines 2 and 3 in Fig. 5.

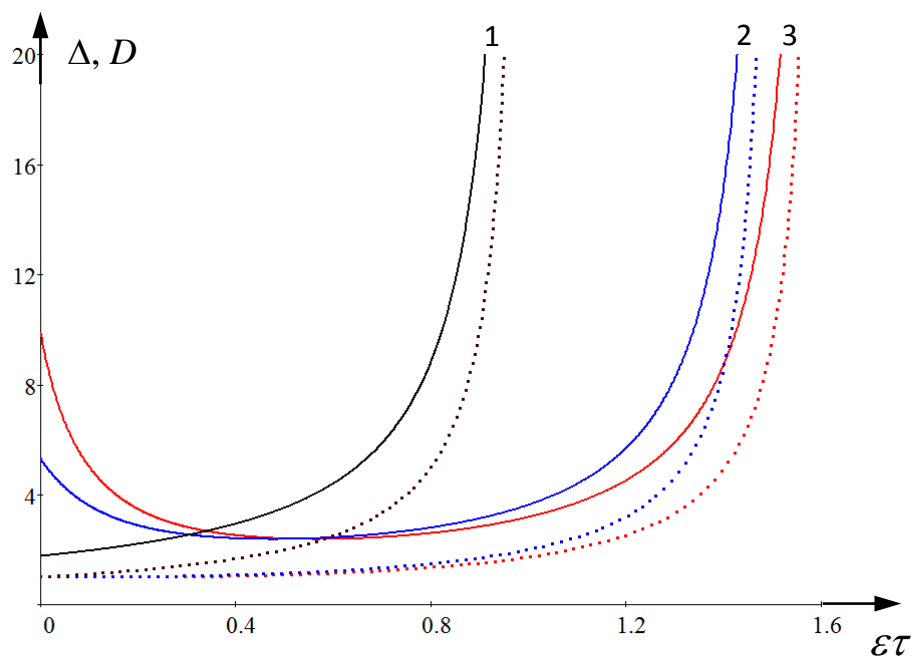

Fig. 5. The soliton front width $\Delta$ (dotted lines) and total width $D$ (solid lines) against time in normalized variables. Lines 1: $B_0 = 0.9999$ (KdV soliton); lines 2: $B_0 = 10^{-2}$ ("fat soliton"); lines 3: $B_0 = 10^{-4}$ (table-top soliton).

As has been mentioned above, the process of soliton evolution under the influence of Earth' rotation is not completed with the terminal decay, but looks more complicated. Numerical calculations show that when the leading soliton decays, it produces an intense trailing wave which in turn evolves into another solitary wave of smaller amplitude (Grimshaw et al., 1998b; Helfrich, 2007). This secondary soliton is accompanied by generation of a trailing wave again. Then the secondary soliton gradually decays generating in turn a trailing wave and new solitary wave. Such quasi-recurrence may occur several times until eventually an envelope solitary wave forms. In the case of relatively small amplitude of initial KdV soliton, the envelope solitary wave can be described by the nonlinear Schrödinger (NLS) equation, whereas when the amplitude of initial solitary wave is big enough, the envelope solitary wave consists of solitary-like waves that propagate through the envelope having a parabolic shape (Helfrich, 2007; Grimshaw & Helfrich, 2008; 2012; Grimshaw et al., 2013; 2016; Whitfield & Johnson, 2014; 2015).

**2.2 The numerical results**

The process of soliton decay under the influence of large-scale Coriolis dispersion was numerically investigated with the help of numerical scheme described in details in (Obregon & Stepanyants, 2012). In the result of this study it was discovered that the adiabatic theory works quite well for small-amplitude solitons and small parameter $\varepsilon$ [see also (Grimshaw et al., 2016)]. This is illustrated by triangles scattered around line 1 in Fig. 2, which was obtained in the direct numerical calculations within the framework of the Ostrovsky Eq. (2.1) with $\mu = 0$, Ur = 12 and $\varepsilon = 3.464 \cdot 10^{-4}$. Small deviations of data points from the theoretical line 1 occur when soliton amplitude becomes small. This can be explained by the influence of non-soliton trailing perturbations which appear gradually in front of propagating soliton due to the periodical boundary condition used in the numerical calculations. Another reason for the deviation of data points from the theoretical line may be caused by the non-adiabatic decay of a solitary wave when its amplitude becomes very small and its shape deviates from the KdV soliton.

However, the adiabatic theory fails to describe the time dependence of amplitude for the "fat" and table-top solitons. It is astonishing that for such solitons of large amplitude numerical data follows the KdV asymptotic dependence (line 1 in Fig. 2) rather than the predicted dependence for the corresponding values of $B_0$. As one can see in Fig. 2, all numerical data obtained both for $B_0 \to 1$ and $B_0 \to 0$ are grouping around line 1. In that figure plusses represent the numerical data for the moderate-amplitude Gardner soliton with $B_0 = 0.134$, $\mu = -0.866$ and Ur = 21.164; diamonds (almost invisible in Fig. 2) represent numerical data for the large-amplitude "fat soliton" with $B_0 = 0.01$, $\mu = -0.99$ and Ur = 23.762; and dots represent numerical data for the large-amplitude table-top soliton" with $B_0 = 4.472 \cdot 10^{-4}$, $\mu = -0.999553$ and, Ur = 23.989.

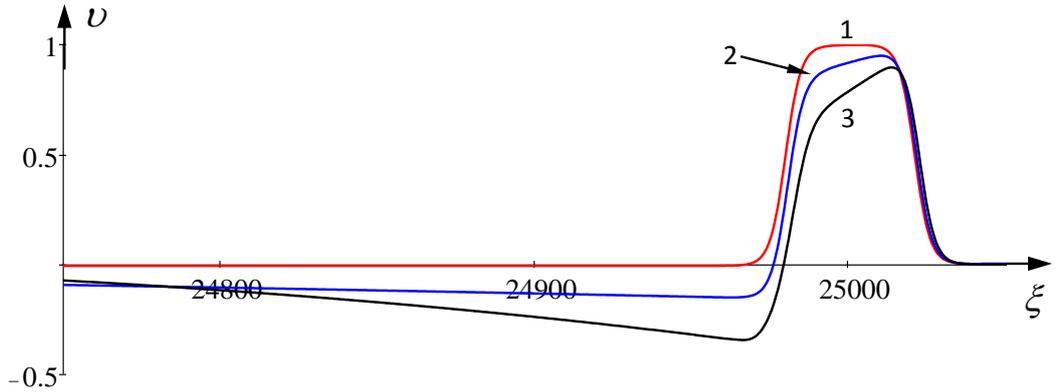

Fig. 6. Table-top soliton at the initial stage of its evolution under the action of large-scale Coriolis dispersion with $\varepsilon = 4.9 \cdot 10^{-3}$. Line 1 – initial Gardner soliton with $B_0 = 4.472 \cdot 10^{-4}$; lines 2 and 3 are the results of its short-time evolution at the consecutive instants of time. The graphics are presented in the limited interval of $\xi$-axis around the initial soliton position, whereas the total spatial interval of calculation was $L = 50{,}000$.

The failure of the adiabatic theory in application to large-amplitude Gardner solitons is related with the composite structure of such solitons. As has been mentioned above, they can be treated as a superposition of coupled kink and anti-kink described by Eq. (2.2b). The kink propagating to the right, generates the negative near-field pedestal which affects the following anti-kink. In the result of that the table-top soliton becomes wry rather than symmetric. This effect is illustrated by Fig. 6, where one can see the initial Gardner soliton of symmetric table-top profile (line 1) and asymmetric pulses accompanied by negative polarity trailing perturbations (lines 2 and 3). Similar results were obtained for the GO equation derived for the electric transmission line (Obregon et al., 2015). Further, in the course of propagation, the leading pulse transfers into a bell-shaped solitary wave resembling a KdV soliton. However, time dependence of the solitary-wave amplitude is described well by the KdV adiabatic theory from the very beginning of evolution.

The advanced asymptotic theory for the description of table-top solitons within the framework of perturbed Gardner equation was developed in (Gorshkov et al., 2012). The proposed approximate approach of soliton evolution was based just on representation of table-top solitons as the compound formations consisting of kinks and antikinks. The obtained theoretical results were in a good agreement with the numerical data published in Ref. (Nakoulima et al., 2004).

In spite of the near-field trailing perturbation is negative, the total mass of generated wave train is positive. This can be shown on the basis of the following reasoning similar to that used in the paper (Grimshaw et al., 2003). The mass of a perturbation can be defined as $M = \int \upsilon(\tau, \xi) d\xi$; within the framework of GO equation this quantity is zero. This follows directly from Eq. (1.1) or in the normalised variables, from Eq. (1.5). If a soliton of nonzero mass is studied in the finite spatial interval of length $L$ (which is the case in the numerical simulations), then the constant pedestal should be added to make the total mass equal to zero: $\upsilon(\tau, \xi) = \upsilon_s(\tau, \xi) - d$, where $\upsilon_s(\tau, \xi)$ is the Gardner soliton described, for instance, by Eq. (2.2a), and the pedestal $d = M_s(0)/L$, where $M_s$ is the soliton mass at the initial instant of time.

As shown below, within the framework of GO equation the mass of a soliton decreases in the course of propagation due to decreasing of soliton amplitude caused by the large-scale dispersion. But the total zero mass of perturbation is conserved within the GO equation, i.e.

$M_s(\tau) + M_{wt}(\tau) - dL = 0$, where $M_{wt}(\tau)$ is the mass of generated wave train. From this equation we find $M_{wt}(\tau) = dL - M_s(\tau) = M_s(0) - M_s(\tau)$ – the result does not depend of $L$. The soliton mass can be readily calculated for the Gardner soliton (2.2a):

$$M_s(\tau) = -\sqrt{\frac{1+B_0}{1-B_0}} \ln\left[1 - \frac{2\sqrt{1-B(\tau)}}{\sqrt{1+B(\tau)} + \sqrt{1-B(\tau)}}\right]. \tag{2.19}$$

Using the solution for the parameter $B(\tau)$ we can present graphically the time dependence of soliton mass and mass of generated wave train, see Fig. 7. As one can see, both of them are positive, but vary in opposite direction: $M_s$ decreasing with time, whereas $M_{wt}$ increasing with time.

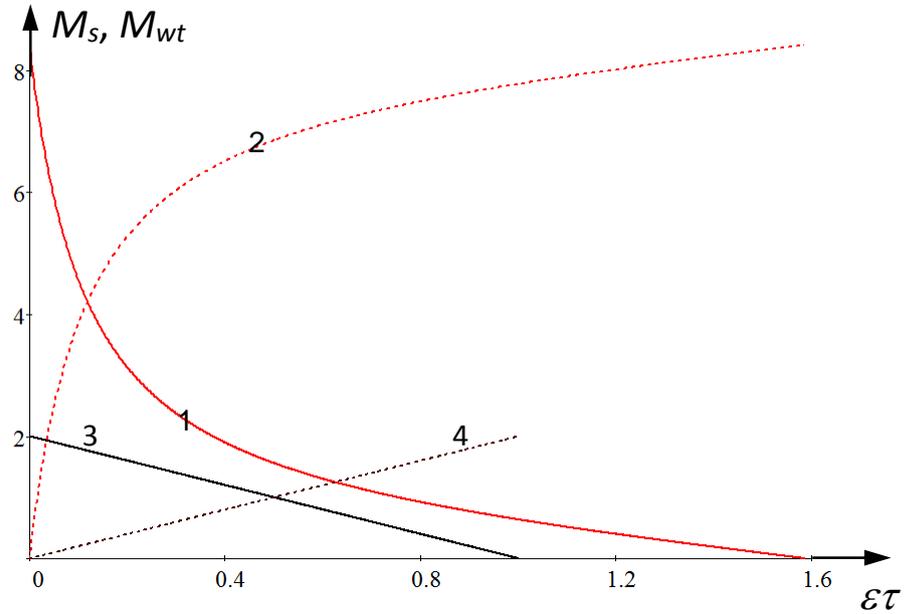

Fig. 7. Time dependence of soliton and wave train masses. Line 1 – table-top soliton with $B_0 = 4.472 \cdot 10^{-4}$ and $\varepsilon = 4.9 \cdot 10^{-3}$; line 2 – the total mass of generated wave train behind the soliton; line 3 – KdV soliton with $B_0 = 1$ and $\varepsilon = 3.464 \cdot 10^{-4}$; line 4 – the total mass of generated wave train behind this soliton.

In the KdV limit, $B_0 \to 1$, Eq. (2.19) reduces to the linear dependence $M_s = 2(1 - \varepsilon\tau)$. The mass of generated wave train in this case is described by simple formula $M_{wt} = 2\varepsilon\tau$. These dependences correspond to lines 3 and 4 in Fig. 7.

## 3. Influence of rotation on the dynamics of bell-shaped solitons when $\alpha_1 > 0$

### 3.1 The theoretical analysis

In this section we consider another case of GO equation (1.1) when the cubic nonlinear coefficient $\alpha_1$ is positive. As has been shown in many papers, there are real oceanic situations when internal waves are described by the Gardner equation with the positive parameter $\alpha_1$ [see, e.g., (Grimshaw et al., 1997; Talipova et al., 1999)]. Such equation also possesses soliton solutions with different profiles and properties in comparison with those studied in Section 2. To study the influence of Earth's rotation on the dynamics of such solitons, we can use again the GO equation in the normalised form (2.1) with $\mu > 0$.

Soliton solution to Eq. (2.1) with $\varepsilon = 0$ and $\mu > 0$ can be described by the same equation (2.2a) where now $B^2 > 1$. In fact, in this case we have two families of solitons: one for $B \geq 1$ and another for $B \leq -1$ (in the former case $\alpha U_0 \geq 0$, whereas in the latter case $\alpha U_0 \leq 0$; both these cases can be met in the real oceanographic situations). All relationships between soliton parameters are described by the very same Eq. (2.3a), and the soliton amplitude is determined by the same formula (2.4). Plots of soliton profiles in terms of $\mu \upsilon$ against $\zeta = (U_r/6\mu)^{1/2} \xi$ are shown in Fig. 8 for several values of parameter $B$. When $B \to 1$ being greater than 1, the soliton (2.2a) reduces to the KdV soliton of infinitely small amplitude, which eventually vanishes when $B$ turns to unity. When $B$ increases, the soliton amplitude also increases and becomes narrower (cf. lines 1, 2 and 3 for $B = 1.5$, 2 and 4, respectively).

For the negative $B$ solitons are of a negative polarity. Their amplitudes infinitely increase as $B \to -\infty$ and they become more and more narrow. However, when $B \to -1$ being less than $-1$, solitons do not vanish, but reduce to the algebraic soliton shown by line 5 in Fig. 8. The formula for the algebraic soliton reads:

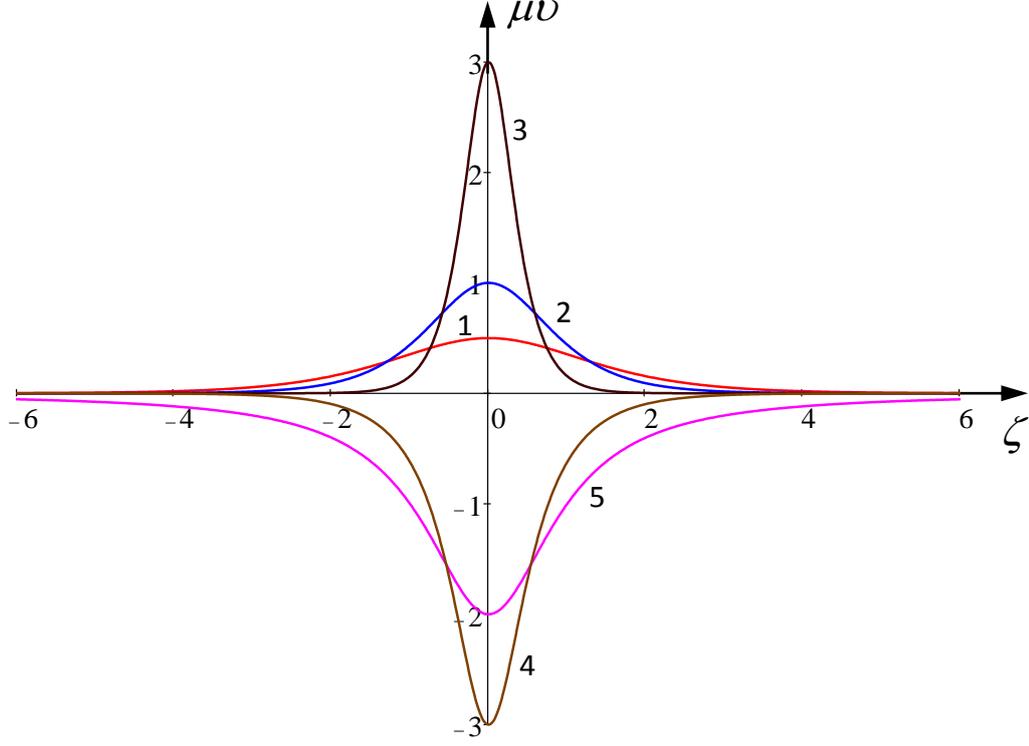

Fig. 8. Normalised Gardner soliton (2.2a) with $\mu > 0$, for several values of parameter $B$. Line 1: $B = 1.5$; line 2: $B = 2$; line 3: $B = 4$; line 4: $B = -2$; line 5: $B = -1$ (the algebraic soliton).

$$\upsilon = \frac{-2}{\mu} \frac{1}{1 + \mathrm{Ur}\xi^2/6\mu}. \tag{3.1}$$

Its amplitude is $-2/\mu$, the characteristic half-width is $(6\mu/\mathrm{Ur})^{1/2}$, and velocity $V = 0$ (in the reference frame moving with the linear velocity $c$).

For further consideration it is convenient to make soliton amplitude $U$ and characteristic width $\Delta$ equal to unity in the dimensionless variables. To this end we put $\mu = B - 1$ and $\mathrm{Ur} = 24/(B + 1)$. Then both families of bell-shaped solitons with $B \geq 1$ and $B \leq -1$ can be presented in the universal form which contains only one free parameter $B$:

$$\upsilon = \frac{1 + B}{1 + B\cosh\left(2\xi - \frac{B+1}{3}\tau\right)}. \tag{3.2}$$

The soliton profiles as described by the dimensionless formula (3.2) are shown in Fig. 9 for several values of parameter $B$ (note that in this variable the algebraic soliton formally has a unit amplitude but zero width).

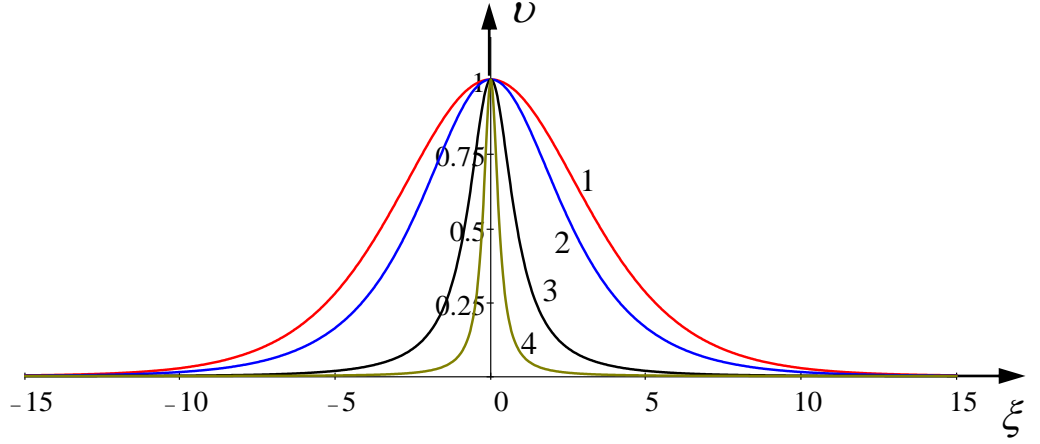

Fig. 9. Normalized Gardner soliton (3.2) for several values of parameter $B$. Line 1: $B = 1$ (KdV soliton); line 2: $B = 10^{10}$ (limiting soliton); line 3: $B = -1.1$; line 4: $B = -1.01$.

When $B$ increases from 1 to infinity, the soliton profile slightly varies from the KdV soliton shown by line 1 in Fig. 9 to the narrower limiting soliton shown by line 2 for $B = 10^{10}$. And when $B$ increases from minus infinity to $-1$, the soliton profile varies from the same limiting soliton shown by line 2 in Fig. 9 to the algebraic soliton of zero width. The approach to the algebraic soliton is shown by lines 3 and 4 for $B = -1.1$ and $-1.01$, correspondingly.

For $\varepsilon \ne 0$ in Eq. (2.1), but sufficiently small, $\varepsilon \ll 1$, we can develop again an asymptotic approach based on the equation of energy balance for a quasi-stationary bell-shaped solitons. By substituting into Eq. (2.6) the soliton solution (2.2a) with $|B| > 1$ and assuming that $B$ is a slowly varying function of time, we ultimately obtain after consecutive integrations (cf. Eq. (2.7) for the "fat" and table-top solitons):

$$\frac{dB}{d\tau} = -8\varepsilon \sqrt{\frac{6\mu}{\mathrm{Ur}}} \frac{B}{\sqrt{B^2 - 1}} \arctan^2 \sqrt{\frac{B-1}{B+1}}. \tag{3.3}$$

In general, an analytical solution to this equation can be presented again in the implicit form in a quadrature:

$$8\sqrt{\frac{6\mu}{\text{Ur}}}\varepsilon\tau = -\int_{B_0}^{B}\frac{\sqrt{B^2-1}}{B\arctan^2\sqrt{\frac{B-1}{B+1}}}dB, \qquad (3.4)$$

where $B_0$ is the initial value of parameter $B$ at $\tau = 0$. However, in the KdV limit ($B_0 \to 1_+$), Eq. (3.4) simplifies and reduces to the equation similar to Eq. (2.9):

$$\frac{dB}{d\tau} = -2\varepsilon\sqrt{(B_0-1)(B-1)}. \qquad (3.5)$$

with the very same exact solution (2.10) for the parameter $B$ or solution (2.11) in terms of soliton amplitude.

In another limit $B_0 \to \pm\infty$ Eq. (3.3) again simplifies and reduces to

$$\frac{dB}{d\tau} = -\varepsilon\pi^2\sqrt{\frac{3\mu}{2\text{Ur}}}\left(1-\frac{4}{\pi B}\right) = -\frac{\varepsilon\pi^2}{4}B_0\left(1-\frac{4}{\pi B}\right). \qquad (3.6)$$

In the last equality it has been used that for the normalized initial soliton with the unit amplitude and characteristic width, the parameters $\mu$ and Ur are (see above): $\mu = B_0 - 1 \approx B_0$ and Ur $= 24/(B_0 + 1) \approx 24/B_0$. Equation (3.6) can be readily integrated resulting in the implicit dependence of $B(\tau)$:

$$\varepsilon\tau = \frac{\pm 4}{\pi^2 B_0}\left(\frac{4}{\pi}\ln\frac{B_0}{B} + B_0 - B\right). \qquad (3.7)$$

Here sign plus (minus) corresponds to the case when $B \to +\infty$ ($B \to -\infty$). This formula can be further simplified for $B \approx B_0$; then we have $B = B_0(1 - \pi^2\varepsilon\tau/4)$, or in terms of soliton amplitude this gives:

$$\frac{U}{U_0} \approx 1 - \frac{\pi^2}{4}\varepsilon\tau. \qquad (3.8)$$

The formulae presented above make sense only until $|B| \geq 1$. When $B > 1$ and decreases reaching 1, the soliton (2.2a) gradually vanishes transforming first into the KdV soliton which completely vanishes in a finite time. Figure 10 illustrates this process in terms of soliton amplitude versus normalized time. Solid lines 1, 2 and 3 in this figure represent numerical solutions of Eq. (3.3) for the different values of parameter $B_0$, and soliton amplitude was further calculated with the help of Eq. (2.4).

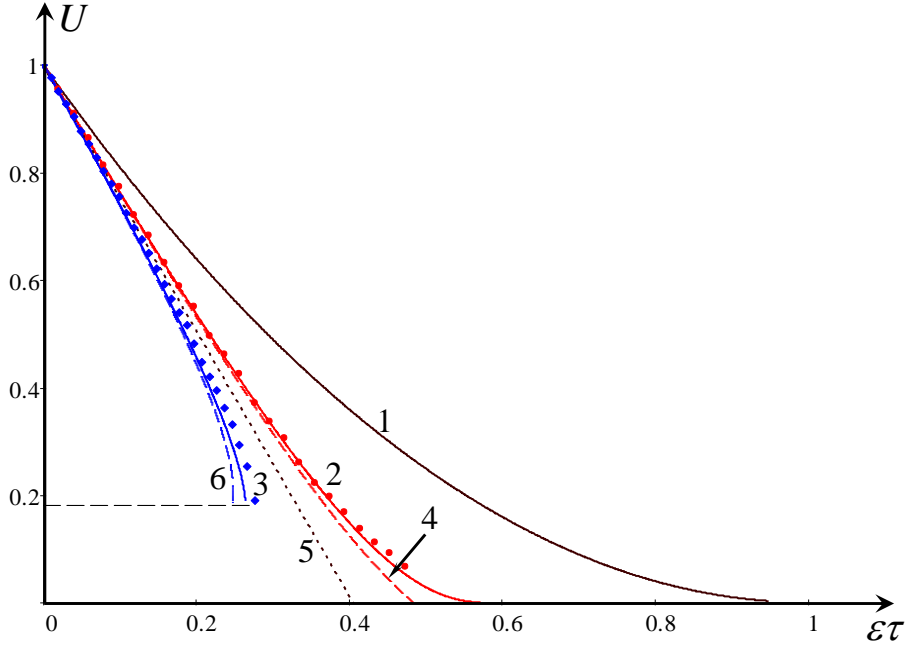

Fig. 10. Bell-shaped soliton amplitude against time in normalized variables. Line 1: $B_0 = 1.01$ (quasi-KdV soliton); line 2: $B_0 = 10$; line 3: $B_0 = -10$. Dashed line 4 represents the asymptotic dependence (3.7) for the same value of $B_0 = 10$. Dotted line 5 displays the limiting case as per Eq. (3.8); and dashed line 6 represents the asymptotic dependence (3.7) for $B_0 = -10$. Symbols present numerical data for the corresponding cases.

The bigger the parameter $B_0$, the faster the soliton decays (cf. lines 1 for $B_0 = 1.01$ and 2 for $B_0 = 10$). The shortest life-time of the bell-shaped solitons with $B_0 > 1$ can be estimated from the asymptotic formula (3.7) which provides a bit underestimated life-time because the formula is formally valid only when soliton amplitude is large enough. According to that formula, soliton amplitude turns to zero at $\tau = \tau_{ext} \equiv 4/(\varepsilon \pi^2)$, which is less than the extinction time for the KdV soliton in $4/\pi^2$ times. The analytical formula for the KdV soliton, Eq. (2.11) with $B_0 = 1$ is indistinguishable from line 1 in Fig. 10.

The situation is different when $B_0 < -1$. In this case the adiabatic theory predicts that the soliton decays until its parameter $B$ increases, but remains less than $-1$. Eventually, when $B$ becomes equal to –1, the soliton transforms into the algebraic soliton (3.2) which represents formally a stationary perturbation moving with the speed of long linear waves *c* in the laboratory coordinate frame. However, as has been shown in Ref. (Pelinovsky & Grimshaw, 1997), the algebraic soliton is structurally unstable; under small perturbations it can be transferred into a mowing breather – a non-stationary solitary wave with the oscillating internal structure (see, e.g., (Grimshaw et al., 2010) and references therein). This is exactly what happens when the Gardner equation is perturbed by the additional term caused by the large-scale Coriolis dispersion. The algebraic soliton loses its identity immediately, and the adiabatic theory become senseless. The numerical solution illustrating the disintegration of algebraic soliton is presented in the next subsection (see Fig. 14 below).

The dotted line 5 in Fig. 10 separates solitons of positive polarity shown in Fig. 8 (their decay lines are shown to the right from the line 5) and solitons of negative polarity whose decay lines are shown to the left from the line 5. All decay lines which correspond to solitons of negative polarity terminate when the parameter *B* becomes equal to –1, and the amplitude becomes equal to $U_{\lim}$ (see horizontal dashed line in Fig. 10), where

$$U_{\lim} = \frac{2}{1-B_0}. \tag{3.9}$$

When Eq. (3.3) is solved for the parameter *B*, all other soliton parameters can be readily obtained as functions of $\tau$, the soliton amplitude $U(\tau)$, velocity $V(\tau)$, and characteristic width $\Delta(\tau)$. The soliton amplitude monotonically decrease with time both for positive and negative initial values of the parameter *B*, although the termination of the adiabatic process is different and depends on the sign of $B_0$. As has been mentioned above, the soliton amplitude formally vanishes at a certain time if $B_0 > 1$. The corresponding extinction time can be found from Eq. (3.4) when *B* turns to ±1; then we have:

$$8\sqrt{\frac{6\mu}{\text{Ur}}}\varepsilon\,\tau_{ext}(B_0) = -\int_{B_0}^{\pm 1} \frac{\sqrt{B^2-1}\,dB}{B\arctan^2\sqrt{\frac{B-1}{B+1}}}. \qquad (3.10)$$

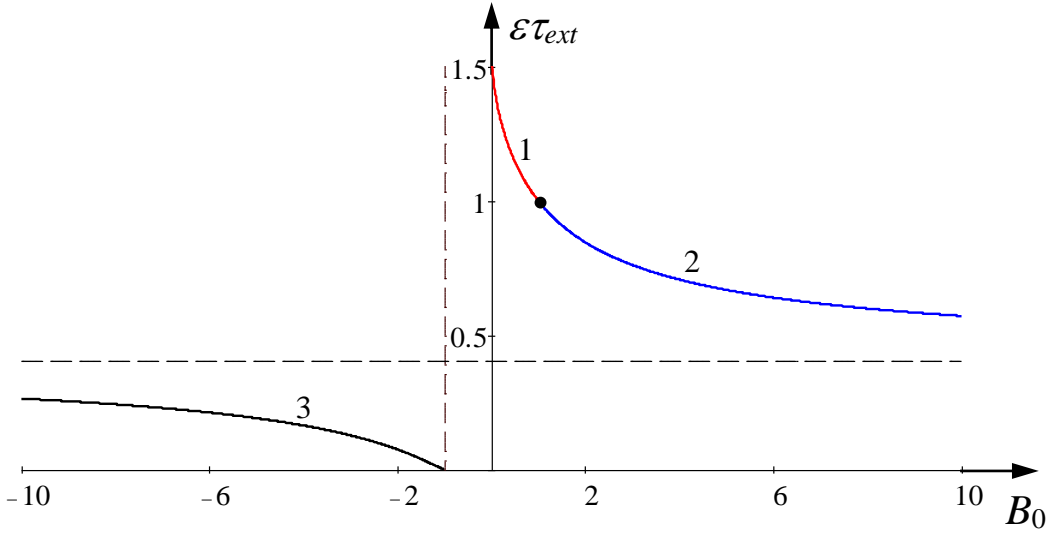

Fig. 11. The extinction time for all types of Gardner solitons against the initial value of parameter $B$. Line 1: table-top and fat solitons when $\mu < 0$; line 2: bell-shaped solitons with $\mu > 0$ and $B_0 > 1$; line 3: bell-shaped solitons with $B_0 < -1$. Black dot corresponds to the KdV soliton, and dashed horizontal line shows the asymptotic value of extinction time $4/\pi^2$ for $B_0 \to \pm\infty$.

Figure 11 shows the dependences of the normalized extinction time on $B_0$ as per Eq. (3.10). In the same figure we present the dependence of extinction time for the table-top solitons as per Eq. (2.15) (see line 1). As follows from this figure, line 2 represents just a continuation of line 1, and the extinction time for the KdV soliton $\varepsilon\tau_{ext} = 1$ exactly corresponds to the point of matching of two branches of one line (see the black point at the border of lines 1 and 2 in Fig. 11). Thus, the adiabatic theory predicts that the extinction time of fat and table-top solitons (when $\mu < 0$) is always greater than the extinction time of KdV soliton, whereas the extinction time of bell-shaped solitons (when $\mu > 0$) is always less than the extinction time of KdV soliton. Further, the characteristic time of bell-shaped solitons with $B_0 < 1$ is always less than the extinction time of bell-shaped solitons with $B_0 > 1$ (cf. lines 2 and 3 in Fig. 11), and the extinction time of an algebraic soliton is formally zero. When $B_0 \to \pm\infty$, the extinction time of bell-shaped solitons

The traversed path for the bell-shaped soliton until its disappearance can be calculated numerically; it is not presented here.

The characteristic soliton scale $\Delta(\tau)$ can be calculated using Eq. (2.3a):

$$\Delta = \sqrt{\frac{24\mu}{\mathrm{Ur}(B^2-1)}} = \sqrt{\frac{B_0^2-1}{B^2-1}}. \tag{3.12}$$

Graphics of $\Delta(\varepsilon t)$ against time are shown in Fig. 13 for different values of $B_0$. In the course of soliton propagation, the width monotonically increases with time and tends to infinity when a soliton with $B_0 > 1$ vanishes or when a soliton with $B_0 < -1$ turns to the algebraic soliton (see, e.g., line 3 in Fig. 13).

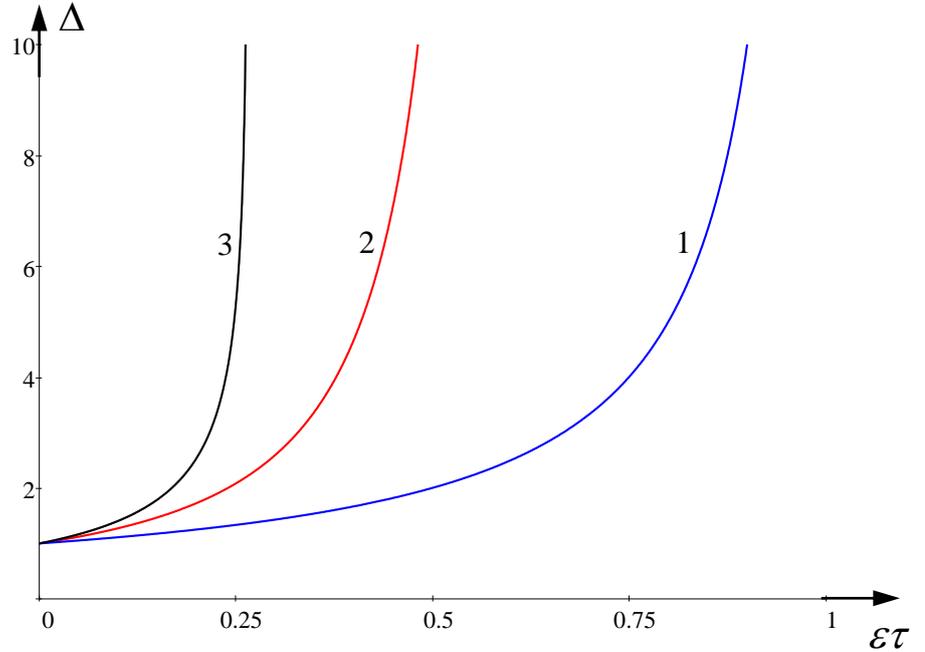

Fig. 13. Characteristic width of a bell-shaped soliton $\Delta$ against time in normalized variables. Line 1: $B_0 = 1.01$ (quasi-KdV soliton); line 2: $B_0 = 10$; line 3: $B_0 = -10$.

### 3.2 The numerical results

The numerical results for soliton decay under the influence of large-scale Coriolis dispersion were obtained with the help of the same numerical code as in Sec. 2.2. The data obtained for the bell-shaped solitons in the case of $\mu > 0$ are in very good agreement with the results of adiabatic

theory – see, for example, symbols next to lines 2 and 3 in Fig. 10. For other cases studied with different values of parameter $B_0$ a similar good agreement between the theory and numerical modelling was obtained.

As has been mentioned above, the extinction time of an algebraic soliton formally is zero (see Fig. 11). Figure 14 illustrates disintegration of algebraic soliton into nonstationary wave trains and, possibly, breathers. As one can see, the soliton radiates a long quasi-sinusoidal wave and quickly changes its own shape and polarity. Then, in the process of evolution it transfers into a nonstationary wave train which, apparently, contains one or two breathers which can represent an intermediate asymptotic of soliton evolution.

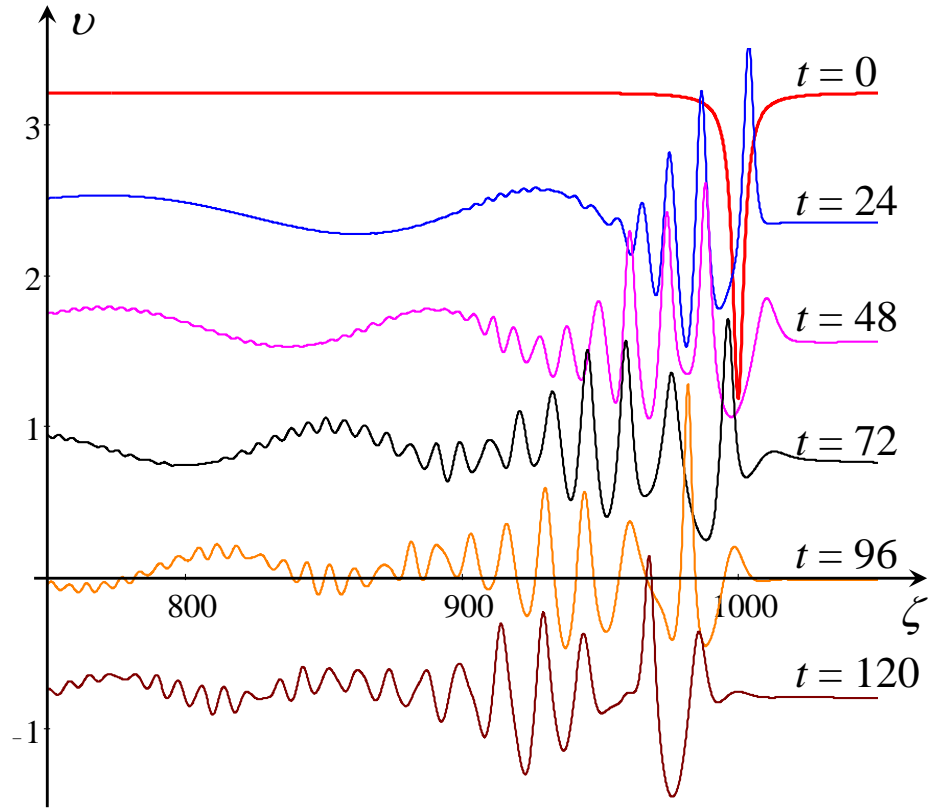

Fig. 14. Disintegration of algebraic soliton into nonstationary wave trains and breathers (fragments).

The breathers of the Gardner equation (Pelinovsky & Grimshaw, 1997), apparently, will decay in turn due to the influence of large-scale Coriolis dispersion. Therefore one can expect again that an envelope NLS-type soliton will be formed eventually after a long-term evolution of an algebraic soliton or a breather. The influence of Earth' rotation on the long-term dynamics of algebraic solitons and breathers will be considered elsewhere.

**4. Conclusion**

In this paper we have analysed the influence of weak Earth' rotation on adiabatic decay of all types of solitary waves within the framework of Gardner–Ostrovsky equation. We consider the equation with both negative and positive coefficient of cubic nonlinear term $\alpha_1$, the term which essentially determines the type of solitary wave. In natural oceanic conditions the GO equation with both signs of this coefficient can occur (Grimshaw et al., 1997; Talipova et al., 1999; Apel et al., 2007).

As has been shown in this paper, Gardner solitons experience slow decay in a weakly rotating fluid due to the radiation of long trailing waves. The adiabatic theory allows us to calculate the decay laws of solitons and estimate the extinction times when solitons completely vanish (in fact they eventually transfer into wave packets similar to envelope solitons of NLS equation (Helfrich, 2007; Grimshaw & Helfrich, 2008; 2012; Grimshaw et al., 2013; 2016; Whitfield & Johnson, 2014; 2015)). The results of direct numerical calculations within the framework of GO equation are in a good agreement with the outcomes of adiabatic theory for bell-shaped solitary waves. In the meantime, the decay law for a table-top soliton of GO equation with $\alpha_1 < 0$ predicted by the adiabatic theory does not agree with the numerical data. The reason of this disagreement is in the composite character of a table-top soliton which consists of a coupled kink and anti-kink. Under the influence of weak rotation the soliton does not keep its symmetric profile, but becomes essentially wry in contradiction with the main assumption of adiabatic theory.

On the basis of the results obtained we have presented estimates for soliton life times in a real ocean.

**Acknowledgement.** Y.S. acknowledges the funding of this study from the State task program in the sphere of scientific activity of the Ministry of Education and Science of the Russian Federation (Project No. 5.1246.2017/4.6) and grant of the President of the Russian Federation for state support of leading scientific schools of the Russian Federation (NSH-2685.2018.5).